\documentclass[aps,prd,onecolumn,10pt,nofootinbib,superscriptaddress,longbibliography]{revtex4-2}

\usepackage[T1]{fontenc}
\usepackage[utf8]{inputenc}
\usepackage{lmodern}
\usepackage{amsmath,amssymb,amsthm,mathtools}
\usepackage[colorlinks=true,linkcolor=blue,citecolor=blue,urlcolor=blue]{hyperref}
\hypersetup{
 pdftitle={Static Einstein--Scalar Reconstruction: Compatibility and Descent},
 pdfauthor={Thomas Schuermann},
 pdfsubject={Static Einstein--scalar reconstruction},
 pdfkeywords={Einstein--scalar systems, inverse problem, compatibility residual, potential descent, regular center}
}

\allowdisplaybreaks

\newtheorem{theorem}{Theorem}[section]
\newtheorem{proposition}[theorem]{Proposition}
\newtheorem{corollary}[theorem]{Corollary}
\theoremstyle{remark}
\newtheorem{remark}[theorem]{Remark}

\newcommand{\dd}{\mathrm{d}}
\newcommand{\ee}{\mathrm{e}}
\newcommand{\cK}{\mathcal{K}}
\newcommand{\cC}{\mathcal{C}}
\newcommand{\cE}{\mathcal{E}}
\newcommand{\cD}{\mathcal{D}}
\newcommand{\Talg}{T^{\mathrm{alg}}}

\begin{document}

\title{Static Einstein--Scalar Reconstruction: Compatibility and Descent}

\author{Thomas Sch\"urmann\\ \textit{D\"usseldorf, Germany}}

\begin{abstract}
Prescribing both the areal-radius metric function $F(r)$ and the scalar profile
$\phi(r)$ generally overdetermines static Einstein--scalar reconstruction. The
temporal and radial Einstein equations determine a redshift function and an
algebraic radial source $V_{\mathrm{alg}}(r)$, but do not by themselves impose
the angular equation or guarantee a single-valued scalar potential. On a
connected static interval we formulate the corresponding membership criterion
for independently prescribed pairs: a potential-independent residual
$\cC[F,\phi]$ must vanish, and $V_{\mathrm{alg}}$ must descend across the full
scalar image to one branch-independent $C^1$ potential, with stationarity at
values attained on constant-profile intervals. The compatibility and descent
requirements are independent. As a positive benchmark, the criterion recovers
the logarithmic Matos--Guzm\'an--N\'u\~nez scalar halo with its exponential
potential. At a radially regular center we compute the leading scalar
backreaction on $F$. Applied to a rational kink profile with rational curvature
interpolation, exact compatibility forces the scalar amplitude to vanish for
every integer $n\geq2$, giving an ansatz-specific compatibility obstruction
rather than a no-hair or classification theorem.
\end{abstract}


\maketitle

\section{Introduction}
\label{sec:introduction}

Static, spherically symmetric configurations of a canonical real scalar field
are a classical testing ground for exact solutions in general relativity. The
massless solution was first obtained by Fisher and later appeared in related
forms in the work of Bergmann and Leipnik, Buchdahl, and Janis, Newman, and
Winicour; Wyman subsequently clarified the static massless solution space
\cite{Fisher1948,BergmannLeipnik1957,Buchdahl1959,JanisNewmanWinicour1968,Wyman1981}.
With self-interaction, the radial geometry, scalar profile, and a single-valued
potential must satisfy a coupled compatibility problem. Closed-form examples
exist \cite{BechmannLechtenfeld1995,MatosGuzmanNunez2000}, while no-hair
theorems and related
obstruction results impose additional restrictions only after regularity,
horizon, sign, and asymptotic hypotheses have been specified
\cite{Chase1970,Bekenstein1972,Bronnikov2001,BronnikovShikin2002}.

When the potential is not fixed in advance, the reduced static
Einstein--scalar equations are commonly organized as inverse problems with one
radial generating function. In a reduced, gauge-fixed, local sense this leaves
one functional freedom, together with integration constants and coordinate
normalizations; boundary conditions, horizons, regularity, and asymptotic
normalizations may then further restrict the family. The present paper does not
use that construction as a solution-generating method. It addresses the more
overdetermined situation in which both the areal-radius metric function $F(r)$
and the scalar profile $\phi(r)$ are prescribed independently on a connected
static interval.

For such data one may use the temporal and radial Einstein equations to
reconstruct the redshift function and a radial quantity
$V_{\mathrm{alg}}(r)$. This step is natural when the starting point is a desired
geometric interpolation, for example one inspired by nonsingular-core metrics
\cite{Dymnikova1992,Hayward2006}. Those metrics are used here only as geometric
motivation; they are not assumed to solve the same canonical Einstein--scalar
theory. The reconstructed quantity $V_{\mathrm{alg}}$ defines an algebraic
radial source, but this source need not be the pullback of a genuine
single-valued potential $V(\phi)$. Moreover, the angular Einstein equation and
the Klein--Gordon equation have not yet been imposed. The problem studied below
is therefore diagnostic rather than generative: it is a membership test for
independently prescribed pairs $(F,\phi)$.

\paragraph*{Main results and scope.}
The paper is local and diagnostic: it tests whether an independently prescribed
pair $(F,\phi)$ belongs, on a connected static interval, to a canonical
Einstein--scalar theory with some single-valued $C^1$ potential. It is not a
global no-hair theorem, a classification of all static solutions, or a
solution-generating inverse construction.

Within this scope, Theorem~\ref{thm:complete-criterion} gives the
necessary-and-sufficient membership criterion. The explicit,
potential-independent residual $\cC[F,\phi]$ must vanish, and the algebraically
reconstructed radial function must descend across the full scalar image to one
branch-independent $C^1$ potential. At every scalar value attained on a
constant-profile interval, the descended derivative must additionally vanish.
Proposition~\ref{prop:compatibility-without-descent} and
Remark~\ref{rem:descent-without-compatibility} show that metric compatibility
and potential descent are logically independent, while
Corollary~\ref{cor:finite-branch-descent} turns descent into a finite-branch
checklist when the scalar has finitely many monotone branches.

At a radially regular center, Theorem~\ref{thm:stationary-center} proves local
scalar constancy at a stationary central value when the force $V_{,\phi}$ is
locally Lipschitz, and Proposition~\ref{prop:backreaction} fixes the first
scalar-dependent coefficient of $F$. The rational kink-type application then
shows that exact compatibility forces the scalar amplitude to vanish for every
integer $n\geq2$. A Matos--Guzm\'an--N\'u\~nez scalar-halo benchmark is
included to show that the same criterion also certifies nonconstant pairs when
both compatibility and descent hold. The contribution is the interval-wise
membership formulation, including the algebraic-source residual,
branch-independent $C^1$ descent, constant-profile stationarity, independence
examples, regular-center coefficient constraint, and rational-family
obstruction.

\paragraph*{Organization of the paper.}
Section~\ref{sec:historical-context} relates this diagnostic problem to standard
one-function inverse constructions and fixes the scope of the contribution. It
is included for orientation; no proof below depends on that comparison.
Section~\ref{sec:compatibility} derives the reduced field equations, introduces
the algebraic source and the compatibility residual, proves the complete local
reconstruction criterion, and records a compact positive benchmark based on the
Matos--Guzm\'an--N\'u\~nez scalar halo. Section~\ref{sec:center} studies the
consequences at a radially regular center. Section~\ref{sec:ansatz} applies the
criterion to the rational
kink-type profile and proves the corresponding compatibility obstruction.
Section~\ref{sec:discussion} summarizes the scope and limitations of the test,
and Section~\ref{sec:conclusions} collects the conclusions. Appendix~\ref{app:series}
gives the coefficient extraction used in the rational obstruction theorem.

We use signature $(-,+,+,+)$, $G>0$, and units $c=\hbar=1$.

\section{Relation to inverse constructions}
\label{sec:historical-context}

When the scalar potential is not fixed in advance, the reduced static
Einstein--scalar equations are often organized as one-function inverse
constructions, with integration constants and coordinate normalizations fixed
later by boundary, regularity, horizon, or asymptotic conditions
\cite{BronnikovShikin2002,TchemarinaTsirulev2009,SolovyevTsirulev2012,Tafel2014,Carloni2014,CadoniEtAl2019}.
The importance of imposing the complete coupled system is illustrated by
Wiltshire's correction of an earlier proposed generator for arbitrary scalar
potentials \cite{Cadoni1991,Wiltshire1992}.

The present paper uses this literature only as a reference point. It does not
choose a generating function and solve for the remaining fields. Instead, both
the areal-radius metric function $F(r)$ and the scalar profile $\phi(r)$ are
prescribed independently on a static interval. The resulting problem is a
membership test: does this overprescribed pair arise from some single-valued
$C^1$ potential in the canonical Einstein--scalar theory?

After translating conventions, the compatibility equation and local quadrature
below are the areal-radius pullbacks of the standard quasiglobal inverse
relations, in particular those used by Tafel. Zeros of $\phi'$, constant-scalar
intervals, continuity of $V_{,\phi}$, and restrictions from nonmonotone scalar
profiles are therefore prior ingredients of the inverse-construction
literature, not separate novelty claims here.

The contribution is the diagnostic packaging for independently prescribed data:
the algebraic-source interpretation of the residual, branch-independent $C^1$
descent across the full scalar image, the constant-profile stationarity test,
the independence examples, the regular-center coefficient constraint, and the
exact obstruction for the rational interpolation family. These local statements
are not intended to replace global inverse constructions.

\section{Complete local reconstruction criterion}
\label{sec:compatibility}

\subsection{Reduced field equations}

Consider
\begin{equation}
 S=\int \dd^4x\,\sqrt{-g}\left[
 \frac{R}{16\pi G}
 -\frac12\nabla_\mu\phi\nabla^\mu\phi
 -V(\phi)
 \right].
 \label{eq:action}
\end{equation}
For the reconstruction statements below, $V\in C^1$ is the standing regularity
assumption. It makes $V_{,\phi}$ continuous and is sufficient for the classical
field equations used here. No $C^2$ regularity of the potential is assumed
unless explicitly stated.

The equations of motion are
\begin{align}
 G_{\mu\nu}&=8\pi G\,T_{\mu\nu},
 \label{eq:einstein}\\
 \Box\phi-V_{,\phi}&=0,
 \label{eq:kg-covariant}
\end{align}
with
\begin{equation}
 T_{\mu\nu}=\nabla_\mu\phi\nabla_\nu\phi
 -g_{\mu\nu}\left[
 \frac12\nabla_\alpha\phi\nabla^\alpha\phi+V(\phi)
 \right].
 \label{eq:stress}
\end{equation}
On a static interval we use areal radius,
\begin{equation}
 \dd s^2=-\ee^{2A(r)}F(r)\,\dd t^2
 +\frac{\dd r^2}{F(r)}+r^2\dd\Omega_2^2,
 \qquad \phi=\phi(r),
 \label{eq:metric}
\end{equation}
with $F>0$. Here $F$ is the radial metric function, while the static lapse is
$N=\ee^A\sqrt{F}$. Unless stated otherwise, the interval is a nonempty open
subset of $(0,\infty)$, $F$ and $\phi$ are $C^2$, and a prime denotes
$\dd/\dd r$. These coordinates are not horizon regular. Some identities may
extend through a regular zero of $F$ by continuity, but every equivalence
statement below is made only on intervals where $F>0$.

\paragraph*{Regularity bookkeeping.}
The assumptions strengthen only where needed. The interval reconstruction
criterion uses $F,\phi\in C^2$ and a descended $V\in C^1$. The local center
uniqueness theorem adds local Lipschitz continuity of $V_{,\phi}$. The regular
center quadrature assumes only one-sided radial $C^2$ regularity at $r=0$; the
leading-coefficient statement assumes the stated asymptotic power law, and the
coefficient filter further assumes one-sided radial analyticity. Finally, the
rational obstruction in Sec.~\ref{sec:ansatz} uses the analyticity of an
explicitly prescribed rational residual on $w>0$. Its small-$w$ coefficient
extractions are algebraic consequences of that rational identity, not additional
assumptions that the spacetime or scalar extend smoothly through $w=0$.

The mixed temporal and radial Einstein equations are
\begin{align}
 -\frac{1-F-rF'}{r^2}
 &=8\pi G\left(-\frac12F\phi'^2-V\right),
 \label{eq:tt}\\
 -\frac{1-F-rF'}{r^2}+\frac{2FA'}{r}
 &=8\pi G\left(\frac12F\phi'^2-V\right).
 \label{eq:rr}
\end{align}
They imply
\begin{align}
 A'&=4\pi G\,r\phi'^2,
 \label{eq:Aprime}\\
 V_{\mathrm{alg}}(r)
 &=\frac{1-F-rF'}{8\pi G\,r^2}
 -\frac12F\phi'^2.
 \label{eq:Valg}
\end{align}
Thus Eqs.~\eqref{eq:tt} and \eqref{eq:rr} determine $A$ up to an additive
constant and determine a radial function $f:=V_{\mathrm{alg}}$. They do not yet
guarantee either the angular Einstein equation or the existence of a genuine
potential $V(\phi)$. Before any descent to a potential is assumed, it is useful
to define the algebraic source
\begin{equation}
 (\Talg)_{\mu\nu}:=\nabla_\mu\phi\nabla_\nu\phi
 -g_{\mu\nu}\left[
 \frac12\nabla_\alpha\phi\nabla^\alpha\phi+f(r)
 \right].
 \label{eq:algebraic-source}
\end{equation}
This tensor is defined by the reconstructed radial function $f$; it is not yet the stress tensor of a scalar theory. With $A$ and $f$ fixed by Eqs.~\eqref{eq:Aprime} and \eqref{eq:Valg}, its temporal and radial components satisfy
\begin{equation}
 G^t{}_t=8\pi G(\Talg)^t{}_t,
 \qquad
 G^r{}_r=8\pi G(\Talg)^r{}_r
 \label{eq:algebraic-tt-rr}
\end{equation}
identically. It becomes the physical tensor \eqref{eq:stress} precisely when $f=V\circ\phi$.

Before any potential descent is assumed, define the potential-independent differential expression
\begin{align}
 \cD[F,\phi]
 &:=\frac{1}{\ee^A r^2}
 \frac{\dd}{\dd r}\left(\ee^A r^2F\phi'\right)
 \label{eq:Dexpression}\\
 &=F\phi''+\left(F'+FA'+\frac{2F}{r}\right)\phi'.
 \label{eq:Dexpanded}
\end{align}
Once a $C^1$ potential is specified on a neighborhood of the scalar image, the radial Klein--Gordon residual is
\begin{equation}
 \cK:=\cD[F,\phi]-V_{,\phi}.
 \label{eq:Kresidual}
\end{equation}

\subsection{Residual identity}

\begin{proposition}[Algebraic-source residual identities and compatibility]
\label{prop:compatibility}
Let $I\subset(0,\infty)$ be a nonempty open connected interval, let $F,\phi\in C^2(I)$, obtain $A$ from Eq.~\eqref{eq:Aprime}, and set $f:=V_{\mathrm{alg}}$. Then the potential-independent identity
\begin{equation}
 \boxed{
 \cC[F,\phi]
 =8\pi G\,r\left(\phi'\cD[F,\phi]-f'\right)
 }
 \label{eq:C-algebraic}
\end{equation}
holds, where
\begin{align}
 \cC[F,\phi]:={}&F''+12\pi G\,r\phi'^2F'\nonumber\\
 &+\left[
 -\frac{2}{r^2}
 +16\pi G\phi'^2
 +16\pi G\,r\phi'\phi''
 +32\pi^2G^2r^2\phi'^4
 \right]F
 +\frac{2}{r^2}.
 \label{eq:compatibility}
\end{align}
After Eqs.~\eqref{eq:tt} and \eqref{eq:rr} have been imposed, one also has the algebraic-source identity
\begin{equation}
 G^\theta{}_{\theta}-8\pi G (\Talg)^\theta{}_{\theta}
 =G^\varphi{}_{\varphi}-8\pi G (\Talg)^\varphi{}_{\varphi}
 =\frac12\cC[F,\phi].
 \label{eq:angular-residual}
\end{equation}
Neither Eq.~\eqref{eq:C-algebraic} nor Eq.~\eqref{eq:angular-residual} assumes that $f$ is a function of $\phi$.

If, in addition, there exist an open interval $U\supseteq\phi(I)$ and a potential $V\in C^1(U)$ such that
\begin{equation}
 f(r)=V(\phi(r)),
 \label{eq:descent}
\end{equation}
then $\Talg=T$, and the chain rule turns Eq.~\eqref{eq:C-algebraic} into
\begin{equation}
 8\pi G\,r\phi'\cK=\cC[F,\phi].
 \label{eq:C-K}
\end{equation}
Consequently, $\cC=0$ is necessary for a full solution. Under the descent assumption, the Klein--Gordon equation implies $\cC=0$ everywhere without division by $\phi'$. Conversely, if $\cC=0$ on $I$, then $\cK=0$ on the set where $\phi'\neq0$ and, by continuity, at every point of $\{\phi'=0\}$ lying in the closure of that set. On an open constant-scalar interval, the additional condition $V_{,\phi}(\phi_0)=0$ must be imposed separately.
\end{proposition}

\begin{proof}
Differentiating Eq.~\eqref{eq:Valg} gives
\begin{equation}
 f'
 =-\frac{F''}{8\pi G\,r}
 -\frac{1-F}{4\pi G\,r^3}
 -\frac12F'\phi'^2-F\phi'\phi''.
 \label{eq:Vprime}
\end{equation}
Using Eq.~\eqref{eq:Dexpanded}, substituting Eq.~\eqref{eq:Aprime}, and combining with Eq.~\eqref{eq:Vprime} gives Eq.~\eqref{eq:C-algebraic}. If Eq.~\eqref{eq:descent} holds, then $f'=V_{,\phi}\phi'$, which yields Eq.~\eqref{eq:C-K}. A direct calculation gives
\begin{equation}
 G^\theta{}_{\theta}
 =\frac12F''+\frac{F'}{r}
 +\frac32A'F'
 +F\left(A''+A'^2+\frac{A'}{r}\right),
 \label{eq:Gtheta}
\end{equation}
from which Eq.~\eqref{eq:angular-residual} follows by using Eqs.~\eqref{eq:Aprime}, \eqref{eq:Valg}, and \eqref{eq:algebraic-source}.
\end{proof}

\begin{proposition}[Exact $C^1$ potential-descent criterion]
\label{prop:descent}
Let $I$ be a nonempty connected interval with nonempty interior, let
$\phi\in C^1(I)$ and $f\in C^1(I)$, and set $E:=\phi(I)$. Here $C^1(I)$ means
continuous on $I$ and $C^1$ on $I^\circ$; if endpoints are included, derivative
identities are imposed on $I^\circ$ and endpoint values are obtained only by the
corresponding one-sided limits. Open intervals $U\supseteq E$ may be bounded,
unbounded, or equal to $\mathbb R$. The following statements are equivalent:
\begin{enumerate}
 \item there exist an open interval $U\supseteq E$ and a function $V\in C^1(U)$ such that $f=V\circ\phi$;
 \item there exist an open interval $U\supseteq E$ and a function $W\in C(U)$ such that
 \begin{equation}
  f'(r)=W(\phi(r))\phi'(r)
  \qquad (r\in I^\circ).
  \label{eq:descent-derivative}
 \end{equation}
\end{enumerate}
When the second statement holds, one may fix $r_0\in I^\circ$ and define
\begin{equation}
 V(y):=f(r_0)+\int_{\phi(r_0)}^{y}W(z)\,\dd z,
 \qquad y\in U.
 \label{eq:descent-construction}
\end{equation}
\end{proposition}

\begin{proof}
Necessity follows from the chain rule with $W=V_{,\phi}$. Conversely,
Eq.~\eqref{eq:descent-construction} defines a function $V\in C^1(U)$ with
$V_{,\phi}=W$. Equation~\eqref{eq:descent-derivative} then shows on $I^\circ$
that
\begin{equation}
 \frac{\dd}{\dd r}\bigl[f(r)-V(\phi(r))\bigr]=0.
\end{equation}
Since $I$ is an interval, $I^\circ$ is connected; the bracket vanishes at $r_0$
and hence on $I^\circ$. If $I$ has included endpoints, continuity extends the
identity to those endpoints.
\end{proof}

\begin{corollary}[Finite-branch functional descent test]
\label{cor:finite-branch-descent}
Assume the hypotheses of Proposition~\ref{prop:descent}. Suppose, in addition,
that the interior critical set $Z:=\{r\in I^\circ:\phi'(r)=0\}$ is a finite union of
points and relatively closed intervals in $I^\circ$, so that $I^\circ\setminus Z$ has finitely many connected components
$I_j$. On each $I_j$ the map $\phi$ is strictly monotone; write
$E_j:=\phi(I_j)$ and let $r_j:E_j\to I_j$ be its inverse. Define
\begin{equation}
 W_j(y):=\frac{f'(r_j(y))}{\phi'(r_j(y))},
 \qquad y\in E_j.
 \label{eq:branch-Wj}
\end{equation}
Then $f$ descends to a $C^1$ potential on an open interval containing
$E=\phi(I)$ if and only if the following finite list of branchwise functional
checks holds. Here ``finite'' refers to the number of monotone branches, branch
pairs, critical endpoint types, and endpoint-to-interior coincidences;
equalities on overlaps are identities on scalar intervals, not merely point
evaluations:
\begin{enumerate}
 \item $f$ is single-valued on scalar fibers,
 \begin{equation}
  \phi(r_1)=\phi(r_2)\quad\Longrightarrow\quad f(r_1)=f(r_2);
 \end{equation}
 \item $f'(r)=0$ at every point of $Z$; critical endpoint values are handled
 by the one-sided endpoint limits in item~4;
 \item the branch derivatives agree on overlaps,
 \begin{equation}
  W_i(y)=W_j(y)\qquad (y\in E_i\cap E_j);
 \end{equation}
 \item at every scalar value $y_0\in E$ reached as an endpoint of one or more
 noncritical branch images $E_j$, all corresponding one-sided limits of
 $W_j(y)$ as $y\to y_0$ exist, are finite, and agree with one another and with
 every interior branch value $W_k(y_0)$ for which $y_0\in E_k$.
\end{enumerate}
In that case the common branch values and matched endpoint limits define a
continuous function $W$ on $E$, which can be extended to a continuous function
on some open interval $U\supseteq E$. Constant-profile intervals impose no
further condition for $C^1$ descent beyond the constancy of $f$; the
Klein--Gordon equation adds the stationarity condition $W(\phi_0)=0$ in
Theorem~\ref{thm:complete-criterion}.
\end{corollary}

\begin{proof}
Necessity follows from Proposition~\ref{prop:descent} and the chain rule. For
sufficiency, the finitely many overlap, endpoint, and endpoint-to-interior
conditions make the functions $W_j$ glue to a single continuous function on the
scalar image away from values attained on constant-profile intervals. At such a
constant value $\phi_0$, the first condition makes $f$ constant on the interval
and the second condition makes Eq.~\eqref{eq:descent-derivative} hold there for
any assigned value of $W(\phi_0)$; if nonconstant branches contain or approach
$\phi_0$, the assigned value is the common branch or limiting value. Because
$E=\phi(I)$ is an interval, this continuous function on $E$ extends to a
continuous function on an open interval $U\supseteq E$.
Proposition~\ref{prop:descent} then gives the required $C^1$ potential.
\end{proof}

\begin{remark}[Branch tests and critical points]
\label{rem:branch-tests}
For a noninjective profile, Proposition~\ref{prop:descent} implies the necessary
branch conditions
\begin{align}
 \phi(r_1)=\phi(r_2)
 &\Longrightarrow
 f(r_1)=f(r_2),
 \label{eq:single-valued}\\
 \phi(r_1)=\phi(r_2),\quad \phi'(r_1)\phi'(r_2)\neq0
 &\Longrightarrow
 \frac{f'(r_1)}{\phi'(r_1)}
 =\frac{f'(r_2)}{\phi'(r_2)}.
 \label{eq:branch-derivative}
\end{align}
Equations~\eqref{eq:single-valued} and \eqref{eq:branch-derivative} are only
pointwise tests. The actual $C^1$ requirement is that $f'/\phi'$ be the pullback
of one continuous function $W$ on an open neighborhood of the full scalar image,
with finite and branch-independent limits at critical scalar values, including
agreement with any interior branch value attained at the same scalar value. The
finite-branch version in Corollary~\ref{cor:finite-branch-descent} is the
practical finite list of functional checks used in examples. On a
constant-profile interval, descent forces $f$ to be constant but leaves
$W(\phi_0)$ undetermined; the Klein--Gordon
equation adds the independent stationarity requirement
$W(\phi_0)=V_{,\phi}(\phi_0)=0$.
\end{remark}

\begin{proposition}[Metric compatibility without potential descent]
\label{prop:compatibility-without-descent}
Fix $r_0>0$, $a\neq0$, and $\phi_0\in\mathbb R$, and write $s:=r-r_0$. Let
\begin{equation}
 \phi(r)=\phi_0+a s^2.
 \label{eq:counterexample-phi}
\end{equation}
There is a unique real-analytic solution of $\cC[F,\phi]=0$ near $r_0$ satisfying
\begin{equation}
 F(r_0)=1,
 \qquad
 F'(r_0)=0.
 \label{eq:counterexample-data}
\end{equation}
On a sufficiently small symmetric interval $I=(r_0-\varepsilon,r_0+\varepsilon)$ with $0<\varepsilon<r_0$, this solution has $F>0$ and $\cC=0$, but $f:=V_{\mathrm{alg}}$ does not descend to a single-valued function of $\phi$.
\end{proposition}

\begin{proof}
For $r_0>0$, the equation $\cC=0$ is a regular linear second-order equation for $F$ with analytic coefficients, so the stated analytic solution exists and is unique. To display the recursion, write $F(r)=\sum_{k\geq0}b_ks^k$. Substitution into $\cC=0$ determines $b_{k+2}$ successively from the lower coefficients. The data \eqref{eq:counterexample-data} give $b_0=1$, $b_1=0$, and coefficient matching yields
\begin{equation}
 b_2=0,
 \qquad
 b_3=-\frac{32\pi G a^2r_0}{3},
 \qquad
 b_4=-\frac{32\pi G a^2}{3},
 \qquad
 b_5=-\frac{16\pi G a^2}{15r_0}.
 \label{eq:counterexample-recursion}
\end{equation}
Consequently,
\begin{align}
 F(r)={}&1-\frac{32\pi G a^2r_0}{3}s^3
 -\frac{32\pi G a^2}{3}s^4
 -\frac{16\pi G a^2}{15r_0}s^5+O(s^6),
 \label{eq:counterexample-F}\\
 f(r)={}&2a^2s^2+\frac{8a^2}{3r_0}s^3
 -\frac{2a^2}{r_0^2}s^4+O(s^5).
 \label{eq:counterexample-f}
\end{align}
Although $\phi(r_0+s)=\phi(r_0-s)$ exactly, Eq.~\eqref{eq:counterexample-f} yields
\begin{equation}
 f(r_0+s)-f(r_0-s)
 =\frac{16a^2}{3r_0}s^3+O(s^5),
\end{equation}
which is nonzero for all sufficiently small nonzero $s$ of fixed sign. Thus the necessary branch condition \eqref{eq:single-valued} fails. This gives an explicit local separation between metric compatibility and potential descent.
\end{proof}

\begin{remark}[Potential descent without metric compatibility]
\label{rem:descent-without-compatibility}
The converse implication also fails. Let $F=1$ and $\phi(r)=\phi_0+qr$ with $q\neq0$ on a bounded open interval $I\subset(0,\infty)$. Then
\begin{equation}
 f=-\frac{q^2}{2}
\end{equation}
descends to the constant potential $V=-q^2/2$, whereas
\begin{equation}
 \cC[F,\phi]
 =16\pi Gq^2+32\pi^2G^2q^4r^2\neq0.
\end{equation}
Thus potential descent does not imply metric compatibility.
\end{remark}

\begin{theorem}[Reconstruction criterion on a static interval]
\label{thm:complete-criterion}
Let $I\subset(0,\infty)$ be a nonempty open connected interval, let $F,\phi\in C^2(I)$ with $F>0$, and define $A$ and $f:=V_{\mathrm{alg}}$ by Eqs.~\eqref{eq:Aprime} and \eqref{eq:Valg}. The following statements are equivalent:
\begin{enumerate}
 \item There exist an open interval $U\supseteq\phi(I)$ and a function $V\in C^1(U)$ for which the metric \eqref{eq:metric}, with the constructed $A$, and the scalar $\phi$ satisfy Eqs.~\eqref{eq:einstein} and \eqref{eq:kg-covariant} on $I$.
 \item The compatibility equation
 \begin{equation}
  \cC[F,\phi]=0
  \qquad\text{on }I
  \label{eq:main-compatibility}
 \end{equation}
 holds, and there exist an open interval $U\supseteq\phi(I)$ and a function $W\in C(U)$ such that
 \begin{equation}
  f'(r)=W(\phi(r))\phi'(r)
  \qquad (r\in I),
  \label{eq:main-descent}
 \end{equation}
 with
 \begin{equation}
  W(\phi_0)=0
  \label{eq:main-constant-condition}
 \end{equation}
 whenever $\phi\equiv\phi_0$ on a nonempty open subinterval of $I$.
\end{enumerate}
When the second statement holds, a potential is obtained from Eq.~\eqref{eq:descent-construction}. Its values outside the scalar image are not fixed by the solution. If $\phi$ has only finitely many monotone branches, the descent part may equivalently be checked by Corollary~\ref{cor:finite-branch-descent}, together with the stationarity condition \eqref{eq:main-constant-condition} on constant-profile intervals.
\end{theorem}

\begin{proof}
Assume the first statement. Since $f=V\circ\phi$, one has $\Talg=T$. The angular Einstein equation and Eq.~\eqref{eq:angular-residual} then give Eq.~\eqref{eq:main-compatibility}. The chain rule gives Eq.~\eqref{eq:main-descent} with $W=V_{,\phi}$, and the Klein--Gordon equation on a constant-scalar interval gives Eq.~\eqref{eq:main-constant-condition}.

Conversely, assume the second statement and construct $V$ from Eq.~\eqref{eq:descent-construction}. Proposition~\ref{prop:descent} gives $f=V\circ\phi$, hence $\Talg=T$, and the mixed temporal and radial Einstein equations hold by construction. Equations~\eqref{eq:main-compatibility} and \eqref{eq:angular-residual} give both angular components by spherical symmetry. On the open set where $\phi'\neq0$, Eq.~\eqref{eq:C-K} gives $\cK=0$. Since $\cK=\cD-W\circ\phi$ is continuous, this equality extends to the boundary of that set. The interior of its complement is a union of constant-profile intervals, on which Eq.~\eqref{eq:main-constant-condition} gives $\cK=-W(\phi_0)=0$. Thus the complete reduced Einstein--Klein--Gordon system holds on $I$.
\end{proof}

\paragraph*{Operational form of the criterion.}
For a prescribed pair $(F,\phi)$ on a connected static interval,
Theorem~\ref{thm:complete-criterion} can be read as the following membership
procedure:
\begin{enumerate}
 \item construct $A$ from Eq.~\eqref{eq:Aprime}; the additive constant only
 rescales the time coordinate;
 \item compute the algebraic source $f=V_{\mathrm{alg}}$ from
 Eq.~\eqref{eq:Valg};
 \item test the potential-independent metric compatibility equation
 $\cC[F,\phi]=0$;
 \item independently test whether $f$ descends to a single-valued $C^1$ function
 of the scalar, equivalently whether there is one continuous branch function $W$
 satisfying Eq.~\eqref{eq:main-descent}; in the finite-branch case use the
 branchwise functional checks of Corollary~\ref{cor:finite-branch-descent};
 \item at each scalar value attained on a nonempty open constant-profile interval,
 impose the stationarity condition $W(\phi_0)=0$.
\end{enumerate}
Failure of any one of these tests rules out the prescribed pair for every
canonical $C^1$ potential on that interval.

\begin{remark}[Spacetime locality versus branch globality]
The theorem is local in spacetime: it assumes neither an asymptotic region nor a
horizon extension. It is nevertheless global over the scalar image traced on
$I$. When $\phi$ is noninjective, the same function $W$ must match every
branch. On a strictly monotone branch, the condition reduces to the familiar
parametric formula $W(\phi(r))=f'(r)/\phi'(r)$.
\end{remark}

\begin{remark}[Role of the Bianchi identity]
After a $C^1$ potential satisfying the descent condition has been fixed, let $\cE^\mu{}_{\nu}:=G^\mu{}_{\nu}-8\pi G T^\mu{}_{\nu}$. Using $\nabla_\mu T^\mu{}_{\nu}=\cK\nabla_\nu\phi$, the radial component of the contracted Bianchi identity gives, once $\cE^t{}_t=\cE^r{}_r=0$,
\begin{equation}
 -\frac{2}{r}\cE^\theta{}_{\theta}
 =-8\pi G\,\phi'\cK.
 \label{eq:Bianchi-residual}
\end{equation}
It relates the two omitted residuals; it does not by itself make either vanish.
Only after all Einstein components hold does the Bianchi identity imply
$\phi'\cK=0$.
\end{remark}

\subsection{Positive benchmark: the Matos--Guzm\'an--N\'u\~nez halo}
\label{subsec:mgn-benchmark}

The criterion is not only obstructive. The constant-radial-coefficient branch of
the Matos--Guzm\'an--N\'u\~nez scalar halo \cite{MatosGuzmanNunez2000} gives a
short nonconstant benchmark.

\begin{proposition}[MGN logarithmic halo]
\label{prop:mgn-benchmark}
Let $I\subset(0,\infty)$ be a nonempty open connected interval. Choose
$r_0>0$, $\phi_0\in\mathbb R$, $0<\beta<1$, and $q\neq0$ with
\begin{equation}
 q^2=\frac{\beta}{4\pi G},
 \label{eq:mgn-qbeta}
\end{equation}
and set
\begin{equation}
 \phi(r)=\phi_0+q\log\frac{r}{r_0},
 \qquad
 F(r)=\frac{1}{1-\beta^2}.
 \label{eq:mgn-pair}
\end{equation}
Then $\cC[F,\phi]=0$, and $V_{\mathrm{alg}}$ descends to the single-valued
potential
\begin{equation}
 V(\varphi)=
 -\frac{\beta}{8\pi G(1-\beta)r_0^2}
 \exp\left[-\frac{2(\varphi-\phi_0)}{q}\right].
 \label{eq:mgn-potential}
\end{equation}
With the time normalization
\begin{equation}
 A(r)=\frac12\log(1-\beta^2)+\beta\log\frac{r}{r_0},
 \label{eq:mgn-A}
\end{equation}
the corresponding line element is
\begin{equation}
 \dd s^2=-\left(\frac{r}{r_0}\right)^{2\beta}\dd t^2
 +(1-\beta^2)\dd r^2+r^2\dd\Omega_2^2.
 \label{eq:mgn-metric}
\end{equation}
Consequently, Eqs.~\eqref{eq:mgn-pair}--\eqref{eq:mgn-metric} solve the full
Einstein--Klein--Gordon system on $I$.
\end{proposition}

\begin{proof}
Equation~\eqref{eq:Aprime} holds because
$A'=\beta/r=4\pi G\,r(q/r)^2$. Since $F$ is constant,
Eq.~\eqref{eq:compatibility} gives
\begin{equation}
 \cC[F,\phi]=\frac{2}{r^2}\left[1-(1-\beta^2)F\right]=0.
 \label{eq:mgn-compatibility}
\end{equation}
The algebraic source is
\begin{equation}
 f(r)=\frac{1-F}{8\pi G r^2}-\frac12F\frac{q^2}{r^2}
 =-\frac{\beta}{8\pi G(1-\beta)}\frac{1}{r^2}.
 \label{eq:mgn-f}
\end{equation}
Because $\phi$ is strictly monotone on $I$,
$r=r_0\exp[(\varphi-\phi_0)/q]$, and Eq.~\eqref{eq:mgn-f} is precisely
$f=V\circ\phi$ with $V$ from Eq.~\eqref{eq:mgn-potential}. There are no
constant-profile intervals, so Theorem~\ref{thm:complete-criterion} gives the
complete field equations. In the notation of Ref.~\cite{MatosGuzmanNunez2000},
this is the constant-radial-coefficient solution with $\kappa_0=8\pi G$ and
rotation-curve exponent $l=2\beta$, after a constant rescaling of $t$.
\end{proof}

\subsection{Factorization and local quadratures}

Using Eq.~\eqref{eq:Aprime}, the compatibility residual can be written as
\begin{equation}
 \cC=F''+3A'F'
 +\left(2A''+\frac{2A'}{r}+2A'^2-\frac{2}{r^2}\right)F
 +\frac{2}{r^2},
 \label{eq:C-A}
\end{equation}
and factorizes as
\begin{equation}
 \cC=
 \left(\frac{\dd}{\dd r}+A'+\frac{2}{r}\right)
 \left(\frac{\dd}{\dd r}+2A'-\frac{2}{r}\right)F
 +\frac{2}{r^2}.
 \label{eq:factorization}
\end{equation}

\begin{proposition}[Areal-radius form of the local compatibility quadrature]
\label{prop:quadrature}
Let $\mathcal{J}\subset(0,\infty)$ be an open interval, let $\phi\in C^2(\mathcal{J})$, choose $r_\star\in\mathcal{J}$, and define $A\in C^2(\mathcal{J})$ by Eq.~\eqref{eq:Aprime}. With
\begin{equation}
 Q(r):=\int_{r_\star}^{r}\ee^{A(s)}\,\dd s,
 \label{eq:Qdef}
\end{equation}
the general $C^2$ solution of $\cC=0$ on $\mathcal{J}$ is
\begin{equation}
 F(r)=\ee^{-2A(r)}r^2
 \left[
 c_2+\int_{r_\star}^{r}
 \frac{\ee^{A(u)}}{u^4}\left(c_1-2Q(u)\right)\dd u
 \right].
 \label{eq:Fquadrature-general}
\end{equation}
This is the complete local solution of the metric compatibility equation for the prescribed profile. It is a compatibility quadrature only: it defines a full reduced Einstein--scalar solution only if the descent and stationary-value conditions of Theorem~\ref{thm:complete-criterion} are also satisfied, and its integration constants are not automatically parameters of one fixed scalar potential.
\end{proposition}

\begin{proof}
Set
\begin{equation}
 Y:=\left(\frac{\dd}{\dd r}+2A'-\frac{2}{r}\right)F.
\end{equation}
Equation~\eqref{eq:factorization} with $\cC=0$ implies
\begin{equation}
 \frac{\dd}{\dd r}\left(\ee^A r^2Y\right)=-2\ee^A,
\end{equation}
so that $Y=\ee^{-A}r^{-2}(c_1-2Q)$. Since
\begin{equation}
 Y=\ee^{-2A}r^2\frac{\dd}{\dd r}
 \left(\ee^{2A}r^{-2}F\right),
\end{equation}
a second integration gives Eq.~\eqref{eq:Fquadrature-general}.
\end{proof}

\begin{remark}[Relation to the quasiglobal inverse construction]
\label{rem:quasiglobal}
On any static interval define
\begin{equation}
 \rho(r):=\rho_\star+\int_{r_\star}^{r}\ee^{A(s)}\,\dd s,
 \qquad
 \mathcal{F}(\rho(r)):=\ee^{2A(r)}F(r).
 \label{eq:quasiglobal-map}
\end{equation}
Since $\dd\rho/\dd r=\ee^A>0$, this is a valid local coordinate transformation and Eq.~\eqref{eq:metric} becomes
\begin{equation}
 \dd s^2=-\mathcal{F}(\rho)\,\dd t^2
 +\frac{\dd\rho^2}{\mathcal{F}(\rho)}
 +r(\rho)^2\dd\Omega_2^2.
 \label{eq:quasiglobal-metric}
\end{equation}
The scalar-gradient relation following from the difference of the Einstein equations, together with the compatibility residual, transforms according to
\begin{align}
 r_{,\rho\rho}&=-4\pi G\,r\phi_{,\rho}^{\,2},
 \label{eq:quasiglobal-scalar}\\
 \left(\frac{\dd\phi}{\dd r}\right)^2
 &=\frac{\dd^2\rho/\dd r^2}{4\pi G\,r\,(\dd\rho/\dd r)},
 \label{eq:inverse-generating-relation}\\
 r^2\mathcal{F}_{,\rho\rho}
 -\mathcal{F}(r^2)_{,\rho\rho}+2
 &=r^2\cC[F,\phi].
 \label{eq:quasiglobal-compatibility}
\end{align}
Consequently, $\cC=0$ has the first integral
\begin{equation}
 r^2\mathcal{F}_{,\rho}
 -\mathcal{F}(r^2)_{,\rho}+2\rho=C_1,
 \label{eq:quasiglobal-first-integral}
\end{equation}
and a second integration gives
\begin{equation}
 \mathcal{F}(\rho)=r(\rho)^2
 \left[
 C_2+\int_{\rho_\star}^{\rho}
 \frac{C_1-2\widetilde\rho}{r(\widetilde\rho)^4}
 \,\dd\widetilde\rho
 \right].
 \label{eq:quasiglobal-quadrature}
\end{equation}
A shift of the origin of $\rho$ is absorbed into $C_1$. Equation~\eqref{eq:Fquadrature-general} is the pullback of Eq.~\eqref{eq:quasiglobal-quadrature}, the local boundary-condition-free form of the inverse quadrature used in Ref.~\cite{Tafel2014}. Here it is retained because it exposes the omitted compatibility equation directly in areal-radius variables and yields the regular-center expansion below.
\end{remark}

For the center analysis, call $r=0$ \emph{radially regular} when
$F,A,\phi\in C^2([0,\varepsilon))$, $F>0$ near the center, and
\begin{equation}
 F(0)=1,
 \qquad
 F'(0)=\phi'(0)=0,
 \qquad
 A(0)\in\mathbb R.
 \label{eq:radial-regularity}
\end{equation}
Equation~\eqref{eq:Aprime} then implies $A'(0)=0$. This is a diagnostic
one-sided radial regularity condition, not a regularity condition on the full
four-dimensional tensor fields in local Cartesian coordinates. A smooth
rotationally invariant scalar has only even powers in its Taylor jet at the
origin; if it is Cartesian analytic, its convergent radial power series is even.
This distinction matters below. The exact rational obstruction is formulated on
$w>0$, whereas a genuinely Cartesian-smooth center would already exclude odd
scalar powers. Under the radial $C^2$ assumptions alone, no expansion beyond
second order is presumed. Accordingly, statements involving odd radial powers
are one-sided radial or interval-wise statements, not assertions of a
Cartesian-smooth scalar core. Any higher-order asymptotic used below is stated
as an additional hypothesis.

Normalize the additive constant in $A$ by
\begin{equation}
 \bar A(r):=A(r)-A(0),
 \qquad
 J(r):=\int_0^r\ee^{\bar A(s)}\,\dd s.
 \label{eq:AbarJ}
\end{equation}

\begin{corollary}[Local regular-center quadrature]
\label{cor:regular-quadrature}
Assume a radially regular center. Every local solution of $\cC=0$ with $F(0)=1$ is
\begin{equation}
 \boxed{
 F_\kappa(r)=\ee^{-2\bar A(r)}
 \left[
 1+\kappa r^2
 -2r^2\int_0^r
 \frac{\ee^{\bar A(u)}J(u)-u}{u^4}\,\dd u
 \right].
 }
 \label{eq:Fquadrature-regular}
\end{equation}
The displayed integrand has a continuous extension to the center with value zero. The constant $\kappa=F''(0)/2$ is the coefficient in $F(r)=1+\kappa r^2+o(r^2)$.
\end{corollary}

\begin{proof}
The argument is first carried out on $(\delta,\varepsilon)$ and only then continued to the center. Set
\begin{equation}
 Y:=\left(\frac{\dd}{\dd r}+2\bar A'-\frac{2}{r}\right)F.
\end{equation}
For $0<\delta<r<\varepsilon$, the first factor in Eq.~\eqref{eq:factorization} gives
\begin{equation}
 \ee^{\bar A(r)}r^2Y(r)
 -\ee^{\bar A(\delta)}\delta^2Y(\delta)
 =-2\int_\delta^r\ee^{\bar A(s)}\,\dd s.
 \label{eq:center-first-integration}
\end{equation}
Radial regularity implies $\phi'(r)=O(r)$, hence $\bar A'(r)=O(r^3)$ and $\bar A(r)=O(r^4)$. It also gives
\begin{equation}
 F(r)=1+\kappa r^2+o(r^2),
 \qquad
 F'(r)=2\kappa r+o(r),
\end{equation}
so that $Y(r)=-2/r+o(r)$. Therefore the second term on the left-hand side of Eq.~\eqref{eq:center-first-integration} tends to zero as $\delta\downarrow0$, and
\begin{equation}
 Y(r)=-2\ee^{-\bar A(r)}r^{-2}J(r).
 \label{eq:center-Y}
\end{equation}
Using
\begin{equation}
 Y=\ee^{-2\bar A}r^2\frac{\dd}{\dd r}
 \left(\ee^{2\bar A}r^{-2}F\right)
\end{equation}
then yields
\begin{equation}
 \frac{\dd}{\dd r}\left(\ee^{2\bar A}r^{-2}F\right)
 =-\frac{2}{r^3}-2h(r),
 \qquad
 h(r):=\frac{\ee^{\bar A(r)}J(r)-r}{r^4}.
 \label{eq:center-h}
\end{equation}
The estimates $\bar A=O(r^4)$ and $J=r+O(r^5)$ imply $h(r)=O(r)$, so $h$ extends continuously by $h(0)=0$. Integrating Eq.~\eqref{eq:center-h} from $\delta$ to $r$ and using
\begin{equation}
 \lim_{\delta\downarrow0}
 \frac{\ee^{2\bar A(\delta)}F(\delta)-1}{\delta^2}
 =\kappa
\end{equation}
gives Eq.~\eqref{eq:Fquadrature-regular} after taking $\delta\downarrow0$.
\end{proof}

At the level of the compatibility equation, $\kappa$ is free. In a fixed scalar
theory it is fixed by the central potential value whenever a full regular
solution exists: taking the center limit in Eq.~\eqref{eq:Valg} gives
\begin{equation}
 V(\phi_0)=-\frac{3\kappa}{8\pi G}.
 \label{eq:kappa-potential}
\end{equation}
Thus Eq.~\eqref{eq:Fquadrature-regular} is the general local regular-center
solution of $\cC=0$, but it is not by itself a one-parameter family of solutions
of one fixed action.

\section{Consequences at a radially regular center}
\label{sec:center}

\subsection{Local scalar constancy at a stationary central value}

\begin{theorem}[Local uniqueness at a stationary central value]
\label{thm:stationary-center}
Let $r=0$ be radially regular, with $\phi(0)=\phi_0$. Suppose that $V\in C^1$ near $\phi_0$, that $V_{,\phi}$ is locally Lipschitz there, and
\begin{equation}
 V_{,\phi}(\phi_0)=0.
 \label{eq:stationary}
\end{equation}
Then every radially regular solution of the Klein--Gordon equation for these
regular coefficients is locally constant:
\begin{equation}
 \phi(r)=\phi_0
\end{equation}
for all sufficiently small $r$. In particular, the conclusion applies to every
full radially regular Einstein--scalar solution satisfying the same hypotheses.
\end{theorem}

\begin{proof}
The Klein--Gordon equation has the integral form
\begin{equation}
 \ee^{A(r)}r^2F(r)\phi'(r)
 =\int_0^r\ee^{A(s)}s^2V_{,\phi}(\phi(s))\,\dd s.
 \label{eq:KG-integral}
\end{equation}
On a sufficiently small interval choose positive constants $m,M$ such that $m\leq\ee^AF$ and $\ee^A\leq M$. Local Lipschitz continuity and Eq.~\eqref{eq:stationary} imply
$|V_{,\phi}(\phi)|\leq L|\phi-\phi_0|$. With
\begin{equation}
 Z(r):=\max_{0\leq s\leq r}|\phi(s)-\phi_0|,
\end{equation}
Eq.~\eqref{eq:KG-integral} gives
\begin{equation}
 |\phi'(r)|\leq \frac{ML}{3m}\,rZ(r),
 \qquad
 Z(r)\leq \frac{ML}{6m}\,r^2Z(r).
\end{equation}
The last coefficient is smaller than one for sufficiently small $r$, so $Z(r)=0$.
\end{proof}

Thus no nonconstant scalar branch can emanate from a radially regular center at
a stationary value when the force $V_{,\phi}$ is locally Lipschitz. The local
constancy propagates through every connected static interval on which the
coefficients remain regular and $F>0$. Indeed, at any $r_0>0$ in such an
interval the Klein--Gordon equation is the regular second-order equation
\begin{equation}
 \phi''=-\left(\frac{F'}{F}+A'+\frac{2}{r}\right)\phi'
 +\frac{V_{,\phi}(\phi)}{F},
 \label{eq:regular-KG-ODE}
\end{equation}
and standard local uniqueness for the data $\phi(r_0)=\phi_0$, $\phi'(r_0)=0$ continues the constant solution. No statement about continuation across a horizon is made here.

For comparison, suppose that a complete local Einstein--Klein--Gordon solution has, for some real $p\geq2$,
\begin{align}
 \phi(r)&=\phi_0+a r^p+o(r^p),
 \qquad a\neq0,
 \label{eq:phi-local}\\
 \phi'(r)&=ap r^{p-1}+o(r^{p-1}),\nonumber\\
 \phi''(r)&=ap(p-1)r^{p-2}+o(r^{p-2}).\nonumber
\end{align}
Since $F=1+O(r^2)$ and $A'=o(r)$, the Klein--Gordon equation yields
\begin{equation}
 V_{,\phi}(\phi(r))
 =a p(p+1)r^{p-2}+o(r^{p-2}).
 \label{eq:Vphi-local}
\end{equation}
For $p=2$, continuity gives $V_{,\phi}(\phi_0)=6a\neq0$. For $p>2$, the limiting value is stationary, but along the branch traced by $r\downarrow0$,
\begin{equation}
 |V_{,\phi}(\phi(r))|
 =p(p+1)|a|^{2/p}
 |\phi(r)-\phi_0|^{1-2/p}\bigl(1+o(1)\bigr).
 \label{eq:Holder}
\end{equation}
Because $0<1-2/p<1$, no locally Lipschitz force with
$V_{,\phi}(\phi_0)=0$ can have this one-sided asymptotic. This is consistent
with Theorem~\ref{thm:stationary-center}: a nonconstant profile with $p>2$
requires weaker regularity of $V_{,\phi}$ at the endpoint value.

\subsection{Leading scalar backreaction}

\begin{proposition}[Leading scalar backreaction]
\label{prop:backreaction}
Assume that $r=0$ is radially regular, that the asymptotics in Eq.~\eqref{eq:phi-local} hold, and that $\cC[F,\phi]=0$ in a neighborhood of the center. Then the radial metric function satisfies
\begin{equation}
 \boxed{
 F(r)=1+\kappa r^2
 -\frac{8\pi G\,a^2p^3}{(p-1)(2p+1)}\,r^{2p}
 +o(r^{2p}).
 }
 \label{eq:backreaction}
\end{equation}
The $r^{2p}$ coefficient is therefore a necessary metric-compatibility condition for the prescribed canonical scalar profile. It does not by itself establish potential descent.
\end{proposition}

\begin{proof}
Set $\beta:=2\pi G a^2p$. Equation~\eqref{eq:Aprime} gives
\begin{equation}
 \bar A(r)=\beta r^{2p}+o(r^{2p}).
 \label{eq:A-local}
\end{equation}
Consequently,
\begin{align}
 \ee^{\bar A(r)}&=1+\beta r^{2p}+o(r^{2p}),\\
 J(r)&=r+\frac{\beta}{2p+1}r^{2p+1}+o(r^{2p+1}),
 \label{eq:J-local}
\end{align}
and the regular integrand in Eq.~\eqref{eq:Fquadrature-regular} has the explicit asymptotics
\begin{equation}
 \frac{\ee^{\bar A(r)}J(r)-r}{r^4}
 =\frac{2\beta(p+1)}{2p+1}r^{2p-3}+o(r^{2p-3}).
 \label{eq:h-local}
\end{equation}
Since $p\geq2$, integration gives
\begin{equation}
 -2r^2\int_0^r
 \frac{\ee^{\bar A(u)}J(u)-u}{u^4}\,\dd u
 =-\frac{2\beta(p+1)}{(p-1)(2p+1)}r^{2p}+o(r^{2p}).
 \label{eq:quadrature-local-term}
\end{equation}
Moreover $\ee^{-2\bar A}=1-2\beta r^{2p}+o(r^{2p})$. Substitution into Eq.~\eqref{eq:Fquadrature-regular} therefore yields
\begin{align}
 F(r)={}&1+\kappa r^2
 -2\beta\left[1+\frac{p+1}{(p-1)(2p+1)}\right]r^{2p}
 +o(r^{2p})\nonumber\\
={}&1+\kappa r^2
 -\frac{8\pi G a^2p^3}{(p-1)(2p+1)}r^{2p}
 +o(r^{2p}),
\end{align}
which proves Eq.~\eqref{eq:backreaction}.
\end{proof}

\begin{corollary}[Local coefficient obstruction]
\label{cor:local-coefficient-obstruction}
Under the hypotheses of Proposition~\ref{prop:backreaction}, suppose a candidate metric has
\begin{equation}
 F(r)=1+\kappa r^2+c_{2p}r^{2p}+o(r^{2p}).
\end{equation}
Compatibility is possible only if
\begin{equation}
 c_{2p}=-\frac{8\pi G\,a^2p^3}{(p-1)(2p+1)}.
 \label{eq:coefficient-obstruction}
\end{equation}
In particular, a candidate that omits the $r^{2p}$ term or assigns it a different coefficient fails metric compatibility and therefore cannot satisfy the full canonical Einstein--scalar equations on any neighborhood of the center.
\end{corollary}

For a profile $\phi-\phi_0\sim\Delta\phi\,w^n$ with $w=r/R$,
Eq.~\eqref{eq:backreaction} requires
\begin{equation}
	F(w)=1+\widehat\kappa w^2
	-\frac{2\alpha n^3}{(n-1)(2n+1)}w^{2n}
	+o(w^{2n}),
	\qquad
	\widehat\kappa:=\kappa R^2,\qquad
	\alpha:=4\pi G(\Delta\phi)^2,
	\label{eq:dimensionless-backreaction}
\end{equation}
whenever $\Delta\phi\neq0$. This provides a local preview of the rational compatibility obstruction below.

\begin{corollary}[Coefficient filter for radial-analytic canonical core data]
\label{cor:analytic-filter}
Within the minimally coupled canonical theory \eqref{eq:action}, let $w=r/R$ and suppose that a radially regular candidate is real analytic as a one-sided function of $w$ and has
\begin{align}
 \phi(w)&=\phi_0+\Delta\phi\,w^n+O(w^{n+1}),
 \qquad n\geq2,\quad \Delta\phi\neq0,\\
 F(w)&=1+\widehat\kappa w^2+\sum_{k=3}^{\infty}c_k w^k,
 \qquad \widehat\kappa:=\kappa R^2,
\end{align}
where $n$ is the first nonzero finite radial order of $\phi-\phi_0$. If $\cC=0$, then
\begin{equation}
 c_k=0\quad(3\leq k<2n),
 \qquad
 c_{2n}=-\frac{2\alpha n^3}{(n-1)(2n+1)},
 \qquad
 \alpha=4\pi G(\Delta\phi)^2.
 \label{eq:analytic-filter}
\end{equation}
Thus any independently prescribed analytic metric interpolation containing a lower-order coefficient $c_k$ with $3\leq k<2n$, or omitting the profile-dependent coefficient at order $w^{2n}$, is locally incompatible with this canonical theory.
\end{corollary}

\begin{proof}
Analyticity supplies the differentiated asymptotics required in Proposition~\ref{prop:backreaction}. Applying that proposition with $p=n$ and $a=\Delta\phi/R^n$ gives
\begin{equation}
 F(w)=1+\widehat\kappa w^2
 -\frac{2\alpha n^3}{(n-1)(2n+1)}w^{2n}
 +o(w^{2n}).
\end{equation}
Comparison with the analytic power series proves Eq.~\eqref{eq:analytic-filter}.
\end{proof}

\begin{remark}[Scope of the radial-analytic coefficient filter]
\label{rem:analytic-filter-scope}
Corollary~\ref{cor:analytic-filter} is local to a radially regular center and assumes the canonical minimally coupled action \eqref{eq:action} and a one-sided radial power series with a first nonzero finite integer order $n$. Cartesian smoothness is stronger and permits only even $n$ for a rotationally invariant scalar. The corollary does not cover smooth but radial-nonanalytic flat profiles such as $\exp(-1/w^2)$, singular centers, phantom kinetic terms, nonminimal couplings, or higher-derivative scalar theories.
\end{remark}

\section{Rational kink-type profile and metric interpolation}
\label{sec:ansatz}

We now test a closed-form family that prescribes both $F$ and $\phi$. The
rational function $x(w)=w^n/(1+w^n)$ interpolates between two scalar values,
while the metric ansatz independently interpolates between two curvature scales
and contains a Schwarzschild-like large-radius term with the same denominator.
For $n\geq3$, the radial center coefficients already exhibit the obstruction:
compatibility removes terms below order $w^{2n}$ and then requires a
scalar-dependent coefficient absent from the reduced metric ansatz. The case
$n=2$ is more delicate, because the center coefficient at order $w^4$ can first
be adjusted by $\lambda_1-\lambda_2$ after regularity sets $\mu=0$; the exact
rational identity supplies the additional condition $\lambda_1=\lambda_2$. The
theorem treats all integers $n\geq2$ on open intervals in $w>0$, in the scope
specified in the Introduction. Including odd $n$ is an interval-wise radial
statement, not an assertion of Cartesian-smooth behavior at $w=0$.

\subsection{Candidate reconstruction}

Let
\begin{equation}
 G>0,
 \qquad R>0,
 \qquad
 \mu,\Lambda_i,\phi_1,\Delta\phi\in\mathbb{R},
 \qquad
 n\in\mathbb{N},\quad n\geq2,
\end{equation}
and define
\begin{equation}
 w:=\frac{r}{R},
 \qquad
 \lambda_i:=\Lambda_iR^2,
 \qquad
 \alpha:=4\pi G(\Delta\phi)^2\geq0.
 \label{eq:dimensionless}
\end{equation}
Here $\mu$ is introduced directly as a dimensionless Schwarzschild-like coefficient in the prescribed radial metric function; no mass interpretation is assumed. Set
\begin{equation}
 x(w):=\frac{w^n}{1+w^n},
 \qquad
 \phi(w):=\phi_1+\Delta\phi\,x(w),
 \qquad
 \phi_2:=\phi_1+\Delta\phi,
 \label{eq:kink}
\end{equation}
with candidate radial metric function
\begin{equation}
 F_n(w):=
 \frac{
 1+w^n-2\mu w^{n-1}
 -\frac{\lambda_1}{3}w^2
 -\frac{\lambda_2}{3}w^{n+2}
 }{1+w^n}.
 \label{eq:Fn}
\end{equation}
For $\Delta\phi\neq0$, the scalar is strictly monotone on $w>0$; hence
$V_{\mathrm{alg}}$ defines a $C^1$ parametric potential on every open subinterval
of that domain. This is only an open-interval statement. It does not by itself
provide a $C^1$ extension of the reconstructed potential to scalar endpoint
values such as $\phi_1$ at $w=0$ or $\phi_2$ at $w=\infty$; the corresponding
endpoint limits must be checked separately by Proposition~\ref{prop:descent} or
Corollary~\ref{cor:finite-branch-descent} whenever such endpoints are included
in the intended scalar domain. Metric compatibility remains an independent
requirement. Equation~\eqref{eq:Aprime} gives
\begin{equation}
 A(w)=A_0+\frac{\alpha n}{2}
 \left(x^2-\frac{2}{3}x^3\right).
 \label{eq:An}
\end{equation}

The small-$w$ expansions explain the later case split:
\begin{align}
 F_2(w)&=1-2\mu w-\frac{\lambda_1}{3}w^2+O(w^3),
 \label{eq:center-n2}\\
 F_3(w)&=1-\left(2\mu+\frac{\lambda_1}{3}\right)w^2+O(w^5),
 \label{eq:center-n3}\\
 F_n(w)&=1-\frac{\lambda_1}{3}w^2-2\mu w^{n-1}+O(w^{n+2}),
 \qquad n\geq4.
 \label{eq:center-ngeneral}
\end{align}
Thus the $n=2$ candidate is not radially regular when $\mu\neq0$, while for $n=3$ the effective central curvature depends on $\lambda_1+6\mu$. Only for $n\geq4$ is the leading central curvature controlled by $\lambda_1$ alone. Odd values of $n$ are retained because the exact theorem is formulated on $w>0$; a Cartesian-smooth scalar center would restrict the family to even $n$.

For later comparison, when $n=2$ and $\mu=0$ one has
\begin{equation}
 F_2(w)=1-\frac{\lambda_1}{3}w^2
 +\frac{\lambda_1-\lambda_2}{3}w^4+O(w^6).
 \label{eq:center-n2-regular}
\end{equation}
The local coefficient filter therefore imposes only
\begin{equation}
 \lambda_1-\lambda_2=-\frac{48}{5}\alpha
 \label{eq:n2-local-relation}
\end{equation}
at this stage; it does not by itself imply $\alpha=0$.

Multiplying Eq.~\eqref{eq:compatibility} by $R^2$ gives
\begin{align}
 \cC_n(w):={}&F_{n,ww}
 +3\alpha w x_w^2F_{n,w}
 \nonumber\\
 &+\left[
 -\frac{2}{w^2}
 +4\alpha x_w^2
 +4\alpha w x_wx_{ww}
 +2\alpha^2w^2x_w^4
 \right]F_n
 +\frac{2}{w^2}.
 \label{eq:Cn}
\end{align}
Metric compatibility requires $\cC_n\equiv0$. This residual condition is an
algebraic statement about the prescribed functions on $w>0$; the theorem below
is deliberately formulated without assuming $F_n>0$. To interpret it as a static
reconstruction obstruction one restricts to connected subintervals on which
$F_n>0$. A full solution must in addition satisfy the descent criterion of
Theorem~\ref{thm:complete-criterion}.

\subsection{Analytic compatibility obstruction}

\begin{theorem}[Compatibility obstruction for the rational family]
\label{thm:rational-obstruction}
Suppose that the residual \eqref{eq:Cn} vanishes on a nonempty open interval in $w>0$. Then
\begin{equation}
 \alpha=0,
 \qquad
 \Delta\phi=0,
 \label{eq:deltaphi-zero}
\end{equation}
and the radial metric function reduces to
\begin{equation}
 F_n(w)=1-\frac{\lambda}{3}w^2,
 \label{eq:maxsym-F}
\end{equation}
with
\begin{equation}
 \begin{array}{lll}
 n=2: & \mu=0, & \lambda_1=\lambda_2=\lambda,\\[1mm]
 n=3: & \lambda_2=\lambda_1+6\mu, & \lambda=\lambda_2,\\[1mm]
 n\geq4: & \mu=0, & \lambda_1=\lambda_2=\lambda.
 \end{array}
 \label{eq:parameter-restrictions}
\end{equation}
For $n=3$, $\mu$ is algebraically redundant. The theorem is a residual-identity
statement; no positivity of $F_n$, horizon extension, or potential descent is
needed for the algebraic conclusion itself.
\end{theorem}

\begin{proof}
Because $1+w^n>0$ for $w>0$, the residual $\cC_n$ is real analytic on that
connected domain. Vanishing on a nonempty open interval therefore makes it an
identity on all of $w>0$. This analytic continuation concerns only the
explicitly prescribed rational function $\cC_n(w)$; it does not assume a static
or regular spacetime continuation through zeros of $F_n$. The subsequent
small-$w$ expansions are therefore coefficient extractions from a rational
identity valid on $w>0$, not additional regularity assumptions at the endpoint
$w=0$. To extract the center coefficients, separate the scalar-free operator
\begin{equation}
 \mathcal{L}_0[g]:=g_{,ww}-\frac{2}{w^2}g,
 \qquad
 \mathcal{L}_0[w^k]=(k-2)(k+1)w^{k-2}.
 \label{eq:L0-monomial}
\end{equation}
The scalar-free part of $\cC_n$ is $\mathcal{L}_0[F_n-1]$. The scalar-dependent terms start at order $w^{2n-2}$ near the center once $F_n=1+O(w)$; the detailed expansions are collected in Appendix~\ref{app:series}.

For $n\geq4$, the term $-2\mu w^{n-1}$ in $F_n-1$ and Eq.~\eqref{eq:L0-monomial} give
\begin{equation}
 \cC_n=-2\mu n(n-3)w^{n-3}+O(w^n),
 \label{eq:C-mu}
\end{equation}
so $\mu=0$. The next relevant term is $[(\lambda_1-\lambda_2)/3]w^{n+2}$, hence
\begin{equation}
 \cC_n=\frac{n(n+3)}{3}
 (\lambda_1-\lambda_2)w^n+O(w^{2n-2}),
 \label{eq:C-lambda}
\end{equation}
which forces $\lambda_1=\lambda_2$. Equation~\eqref{eq:Fn} now reduces exactly to Eq.~\eqref{eq:maxsym-F}. Since $\mathcal{L}_0[-\lambda w^2/3]=0$, the leading residual comes from
\begin{equation}
 4\alpha\left(x_w^2+w x_wx_{ww}\right)F_n
 =4\alpha n^3w^{2n-2}+O(w^{2n}),
\end{equation}
and therefore
\begin{equation}
 \cC_n=4\alpha n^3w^{2n-2}+O(w^{2n}).
 \label{eq:C-alpha}
\end{equation}
Thus $\alpha=0$.

For $n=3$, the $w^2$ term in $F_3-1$ lies in the kernel of $\mathcal{L}_0$, while the $w^5$ term produces the next scalar-free contribution. Together with the leading scalar term this gives
\begin{equation}
 \cC_3=
 6(\lambda_1-\lambda_2+6\mu)w^3
 +108\alpha w^4+O(w^6).
 \label{eq:C-n3}
\end{equation}
The independent powers imply $\lambda_2=\lambda_1+6\mu$ and $\alpha=0$. Under the first relation, the numerator of Eq.~\eqref{eq:Fn} is
\begin{align}
 &1+w^3-2\mu w^2
 -\frac{\lambda_1}{3}w^2
 -\frac{\lambda_2}{3}w^5
 \nonumber\\
 &\hspace{2cm}=(1+w^3)
 \left(1-\frac{\lambda_2}{3}w^2\right),
 \label{eq:n3-factorization}
\end{align}
which proves Eq.~\eqref{eq:maxsym-F} and the redundancy of $\mu$.

For $n=2$, Eq.~\eqref{eq:L0-monomial} applied to the linear term $-2\mu w$ gives
\begin{equation}
 \cC_2=\frac{4\mu}{w}+O(w),
 \label{eq:C-n2-center}
\end{equation}
so $\mu=0$. At the level of the center expansion, the next coefficient only gives the temporary relation \eqref{eq:n2-local-relation}; the scalar amplitude is therefore not eliminated locally. The required additional information comes from the exact rational identity. After this reduction, clearing the common denominator in the exact expression for $\cC_2$ gives
\begin{equation}
 \cC_2(w)=\frac{2w^2}{3(1+w^2)^9}\,P(w^2),
 \label{eq:C-n2-polynomial}
\end{equation}
where $P(z)$ is the polynomial displayed in Eq.~\eqref{eq:P-full}. The full polynomial is recorded to make the algebraic common-denominator calculation reproducible; the proof uses just two coefficients,
\begin{equation}
 P(0)=48\alpha+5(\lambda_1-\lambda_2),
 \qquad
 [z^7]P=\lambda_1-\lambda_2.
 \label{eq:C-n2-polynomial-coefficients}
\end{equation}
Since \eqref{eq:C-n2-polynomial} vanishes identically for $w>0$, the polynomial $P$ is identically zero. The two-coefficient extraction is decisive: the $z^7$ coefficient first gives $\lambda_1=\lambda_2$, and the constant coefficient then gives $\alpha=0$. Equation~\eqref{eq:Fn} consequently reduces to Eq.~\eqref{eq:maxsym-F}. Since $G>0$, Eq.~\eqref{eq:dimensionless} then implies $\Delta\phi=0$ in every case.
\end{proof}

For orientation, the coefficient logic of the proof can be summarized as follows.
The entries are functional coefficient identities in the small-$w$ expansion,
except for the last $n=2$ step, which uses the cleared rational polynomial
$P(z)$:
\begin{center}
\small
\renewcommand{\arraystretch}{1.15}%
\begin{tabular}{c|c|c|c}
case & first decisive term & consequence & final step \\ \hline
$n\geq4$ & $-2\mu n(n-3)w^{n-3}$ & $\mu=0$ & $\lambda_1=\lambda_2$, then $\alpha=0$ \\
$n=3$ & \begin{tabular}{@{}c@{}}$6(\lambda_1-\lambda_2+6\mu)w^3$\\$+108\alpha w^4$\end{tabular} & $\lambda_2=\lambda_1+6\mu$ & $w^4$ term gives $\alpha=0$ \\
$n=2$ & \begin{tabular}{@{}c@{}}$4\mu/w$ and the exact\\polynomial $P(z)$\end{tabular} & \begin{tabular}{@{}c@{}}$\mu=0$; then $[z^7]P=0$\\gives $\lambda_1=\lambda_2$\end{tabular} & $P(0)=0$ gives $\alpha=0$
\end{tabular}
\end{center}

\begin{corollary}[Completion as a constant-scalar solution]
\label{cor:constant-completion}
Suppose, in addition, that the ansatz data define a solution of the complete Einstein--Klein--Gordon system for a potential $V\in C^1(U)$ on an open neighborhood $U$ of the constant scalar value $\phi_1$. Then $A$ is constant and, after a constant rescaling of $t$, the surviving line element is the static chart of a four-dimensional maximally symmetric spacetime on each connected interval on which $F>0$. Its metric function is
\begin{equation}
 F(r)=1-\frac{\Lambda}{3}r^2,
 \qquad
 \Lambda:=\frac{\lambda}{R^2}.
 \label{eq:physical-Lambda}
\end{equation}
It is a full constant-scalar solution precisely when
\begin{equation}
 V(\phi_1)=\frac{\lambda}{8\pi G R^2}
 =\frac{\Lambda}{8\pi G},
 \qquad
 V_{,\phi}(\phi_1)=0.
 \label{eq:constant-completion}
\end{equation}
The case $\Lambda=0$ is Minkowski space.
\end{corollary}

\begin{proof}
Theorem~\ref{thm:rational-obstruction} gives $\Delta\phi=0$, so Eq.~\eqref{eq:Aprime} makes $A$ constant; its additive value is removed by rescaling $t$. Equation~\eqref{eq:Valg} evaluated on Eq.~\eqref{eq:physical-Lambda} gives the first condition in Eq.~\eqref{eq:constant-completion}, while the Klein--Gordon equation for a constant scalar gives the second. Conversely, these two conditions satisfy all reduced field equations.
\end{proof}

As a static reconstruction statement, Theorem~\ref{thm:rational-obstruction}
is applied only on connected subintervals where $F_n>0$. A complete solution
must still satisfy the potential-descent and stationary-value conditions of
Theorem~\ref{thm:complete-criterion}; the theorem proves only the compatibility
part for this prescribed rational family.

\section{Discussion}
\label{sec:discussion}

The sequence of tests is part of the result. Before a matter potential has been
constructed,
\begin{equation}
 \cC=8\pi G r\left(\phi'\cD-f'\right),
\end{equation}
and $\cC/2$ is the angular Einstein residual relative to the algebraic source
$\Talg$. Only after $f$ descends to a $C^1$ potential does $\Talg$ become the
physical stress tensor and the identity reduce to
$\cC=8\pi G r\phi'\cK$. The Bianchi identity then relates the angular and
Klein--Gordon residuals, but it does not remove the separate checks at critical
points or on constant scalar intervals. This is why the independence examples
are structural rather than ornamental. The Matos--Guzm\'an--N\'u\~nez benchmark
shows the complementary case: when both tests pass, the same formalism certifies
a known nonconstant solution.

The factorized quadrature has the same limited status. It solves the linear
metric-compatibility equation for a prescribed scalar profile and is useful for
local coefficient extraction, but it does not guarantee potential descent. An
integration parameter that remains in the descended function $V(\phi)$ changes
the action when varied and should not automatically be interpreted as a mass,
charge, or thermodynamic integration constant of one fixed theory. At a radially
regular center, compatibility leaves the curvature coefficient $\kappa$ free,
whereas a fixed central potential value determines it through
Eq.~\eqref{eq:kappa-potential}.

The regularity assumptions are deliberately separated. The reconstruction
criterion is a $C^2/C^1$ interval theorem; the center uniqueness theorem adds a
locally Lipschitz force; the coefficient filter assumes one-sided radial
analyticity; and the rational obstruction uses the analyticity of the displayed
residual on $w>0$. In the rational family the center coefficients settle all
cases except the final $n=2$ step, where the exact polynomial identity supplies
$\lambda_1=\lambda_2$ and hence $\alpha=0$. The conclusion therefore rests on
the prescribed residual identity, not on horizon or asymptotic assumptions.\\

As a literature sanity check, the criterion does not exclude known
nonconstant Einstein--scalar solutions. The massless
Fisher/Janis--Newman--Winicour/Wyman family
\cite{Fisher1948,JanisNewmanWinicour1968,Wyman1981} and the logarithmic
Matos--Guzm\'an--N\'u\~nez halo \cite{MatosGuzmanNunez2000} lie within the
canonical static spherical setting and pass the corresponding compatibility
and descent checks after translating conventions. By contrast,
Dymnikova--Hayward-type nonsingular-core metrics
\cite{Dymnikova1992,Hayward2006} are only geometric motivation here, not
canonical one-real-scalar reconstructions. Thus the obstruction derived above
is a diagnostic statement about prescribed pairs $(F,\phi)$, not a no-hair
theorem or a classification result.

\section{Conclusions}
\label{sec:conclusions}

For independently prescribed areal-radius data $(F,\phi)$ on a static interval,
the reduced Einstein--Klein--Gordon system is solved exactly when the metric
residual $\cC[F,\phi]$ vanishes and $V_{\mathrm{alg}}$ descends across the full
scalar image to one branch-independent $C^1$ potential. Values attained on
constant-profile intervals additionally require stationarity. This separates
metric compatibility, potential descent, and constant-profile stationarity; the
Matos--Guzm\'an--N\'u\~nez halo gives a compact nonconstant benchmark passing
all tests.

At a radially regular center, a stationary scalar value with locally Lipschitz
force admits only the locally constant scalar solution. If a nonconstant profile
starts as $\phi-\phi_0\sim ar^p$, metric compatibility fixes the first
scalar-dependent term of $F$ at order $r^{2p}$. This condition is necessary but
not a substitute for the full descent criterion.

For the rational kink-type scalar profile combined with the rational
curvature-interpolation metric, the exact residual identity forces
$\Delta\phi=0$ for every integer $n\geq2$. The family therefore collapses to the
constant-scalar branch, with
$F(r)=1-\Lambda r^2/3$ on each connected static patch. This remaining branch
solves the complete system precisely when the constant scalar is a stationary
point of a potential with value $\Lambda/(8\pi G)$. The obstruction is
ansatz-specific and is neither a no-hair theorem nor a classification result.

\section*{Declarations}
The author declares no competing interests. No data were generated or analyzed for this work.

\appendix

\section{Coefficient extraction in the rational obstruction theorem}
\label{app:series}

For $w\to0$,
\begin{align}
 x_w&=nw^{n-1}-2nw^{2n-1}+O(w^{3n-1}),
 \label{eq:series-xw}\\
 x_{ww}&=n(n-1)w^{n-2}
 -2n(2n-1)w^{2n-2}+O(w^{3n-2}).
 \label{eq:series-xww}
\end{align}
It is useful to write the scalar-free part of Eq.~\eqref{eq:Cn} as
\begin{equation}
 F_{n,ww}-\frac{2}{w^2}F_n+\frac{2}{w^2}
 =\mathcal{L}_0[F_n-1],
 \qquad
 \mathcal{L}_0[w^k]=(k-2)(k+1)w^{k-2}.
 \label{eq:appendix-L0}
\end{equation}
Thus the $w^2$ curvature term lies in the kernel of $\mathcal{L}_0$, while every other monomial produces a coefficient immediately.

For $n\geq4$,
\begin{equation}
 F_n=1-\frac{\lambda_1}{3}w^2
 -2\mu w^{n-1}
 +\frac{\lambda_1-\lambda_2}{3}w^{n+2}
 +O(w^{2n-1}).
 \label{eq:series-F-general}
\end{equation}
Equation~\eqref{eq:appendix-L0} gives
\begin{align}
 \mathcal{L}_0[-2\mu w^{n-1}]
 &=-2\mu n(n-3)w^{n-3},
 \label{eq:appendix-mu}\\
 \mathcal{L}_0\left[\frac{\lambda_1-\lambda_2}{3}w^{n+2}\right]
 &=\frac{n(n+3)}{3}(\lambda_1-\lambda_2)w^n.
 \label{eq:appendix-lambda}
\end{align}
All scalar-dependent terms begin at order $w^{2n-2}$, so Eqs.~\eqref{eq:appendix-mu} and \eqref{eq:appendix-lambda} yield Eqs.~\eqref{eq:C-mu} and \eqref{eq:C-lambda} successively. After $\mu=0$ and $\lambda_1=\lambda_2$, one has $F_n=1-\lambda_1w^2/3$ exactly. The only contribution at order $w^{2n-2}$ is
\begin{align}
 4\alpha\left(x_w^2+w x_wx_{ww}\right)F_n
 &=4\alpha\left[n^2+n^2(n-1)\right]w^{2n-2}
 +O(w^{2n})
 \nonumber\\
 &=4\alpha n^3w^{2n-2}+O(w^{2n}),
 \label{eq:appendix-alpha}
\end{align}
which is Eq.~\eqref{eq:C-alpha}. The term $3\alpha w x_w^2F_{n,w}$ is $O(w^{2n})$, and the $\alpha^2$ term is of still higher order.

For $n=3$,
\begin{equation}
 F_3=1-\left(2\mu+\frac{\lambda_1}{3}\right)w^2
 +\left(2\mu+\frac{\lambda_1-\lambda_2}{3}\right)w^5
 +O(w^8).
 \label{eq:series-F3}
\end{equation}
Because $\mathcal{L}_0[w^2]=0$ and $\mathcal{L}_0[w^5]=18w^3$, the scalar-free coefficient is
\begin{equation}
 18\left(2\mu+\frac{\lambda_1-\lambda_2}{3}\right)w^3
 =6(\lambda_1-\lambda_2+6\mu)w^3.
\end{equation}
Equation~\eqref{eq:appendix-alpha} with $n=3$ supplies the independent term $108\alpha w^4$, proving Eq.~\eqref{eq:C-n3}.

For $n=2$,
\begin{equation}
 F_2=1-2\mu w-\frac{\lambda_1}{3}w^2
 +2\mu w^3+\frac{\lambda_1-\lambda_2}{3}w^4
 +O(w^5).
 \label{eq:series-F2}
\end{equation}
The linear term gives
\begin{equation}
 \mathcal{L}_0[-2\mu w]=\frac{4\mu}{w},
\end{equation}
which proves Eq.~\eqref{eq:C-n2-center} and forces $\mu=0$. With $\mu=0$, the center expansion of the residual is
\begin{equation}
 \cC_2(w)=
 \left[
 32\alpha+\frac{10}{3}(\lambda_1-\lambda_2)
 \right]w^2+O(w^4).
 \label{eq:C-n2-local}
\end{equation}
Its leading coefficient is equivalent to the local relation \eqref{eq:n2-local-relation}; this is why the center expansion alone does not yet rule out $\alpha\neq0$.

For the exact identity, direct differentiation and collection over the common denominator $3(1+w^2)^9$ give
\begin{equation}
 \cC_2(w)=\frac{2w^2}{3(1+w^2)^9}P(w^2).
\end{equation}
This finite polynomial identity is included to make the $n=2$ step reproducible. Writing $\delta_\lambda:=\lambda_1-\lambda_2$, the numerator polynomial is
\begin{align}
 P(z)={}&\delta_\lambda z^7
 +\left(4\alpha\lambda_2+11\delta_\lambda\right)z^6
 \nonumber\\
 &+\left[4\alpha(4\lambda_1-7\lambda_2-12)
 +45\delta_\lambda\right]z^5
 \nonumber\\
 &+\left[-16\alpha^2\lambda_2
 +4\alpha(5\lambda_1-27\lambda_2-36)
 +95\delta_\lambda\right]z^4
 \nonumber\\
 &+\left[16\alpha^2(3-\lambda_1)
 -4\alpha(9\lambda_1+29\lambda_2+24)
 +115\delta_\lambda\right]z^3
 \nonumber\\
 &+\left[48\alpha^2
 -4\alpha(17\lambda_1+10\lambda_2-24)
 +81\delta_\lambda\right]z^2
 \nonumber\\
 &+\left[4\alpha(36-7\lambda_1)
 +31\delta_\lambda\right]z
 +48\alpha+5\delta_\lambda.
 \label{eq:P-full}
\end{align}
The proof of Theorem~\ref{thm:rational-obstruction} uses only the two endpoint coefficients of this displayed numerator: the highest coefficient $[z^7]P$ and the constant coefficient $P(0)$, as recorded in Eqs.~\eqref{eq:C-n2-polynomial-coefficients}. Thus no factorization or root analysis of $P$ is required.

\end{document}